\documentclass[twocolumn,aps,prb,reprint,noshowpacs,letterpaper,floatfix]%
{revtex4-1}
\usepackage{amsmath,amssymb}
\usepackage{txfonts}
\usepackage{bm}
\usepackage{textcomp}
\usepackage[pdftex]{graphicx}
\usepackage[bookmarks=false]{hyperref}
\usepackage{graphicx}
\usepackage{dcolumn}
\usepackage{bm}
\usepackage{color}

\begin{document}


\title{Indication of Ferromagnetic Quantum Critical Point in Kondo Lattice CeRh$_6$Ge$_4$}

\author{Hisashi Kotegawa, Eiichi Matsuoka, Toshiaki Uga, Masaki Takemura, Masahiro Manago, \\
Noriyasu Chikuchi, Hitoshi Sugawara, Hideki Tou, Hisatomo Harima}

\affiliation{
Department of Physics, Kobe University, Kobe 657-8501, Japan
}

\date{\today}

\begin{abstract}
We report resistivity measurements under pressure for Kondo-lattice ferromagnet CeRh$_6$Ge$_4$, and present that a quantum ferromagnetic (FM) phase transition is easily achieved.
In most clean metallic ferromagnets, a quantum critical point (QCP) at zero field is avoided by changing the FM transition to a discontinuous transition or to an antiferromagnetic transition. 
In CeRh$_6$Ge$_4$, to the contrary, the Curie temperature of 2.5 K decreases continuously as increasing pressure without any clear signature that the transition changes to first order. 
The obvious non Fermi liquid behavior is observed in the vicinity of the quantum FM phase transition.
The experimental data do not contradict a picture in which CeRh$_6$Ge$_4$ shows the FM QCP at zero field.
Band structure calculation suggests the unusual electronic state of CeRh$_6$Ge$_4$ among Ce-based Kondo lattices.
CeRh$_6$Ge$_4$ deserves further investigations and will be a key material to understand the matter of the FM QCP.

\end{abstract}

\maketitle

A quantum critical point (QCP) is a phase transition at zero temperature accompanied by a continuous change of the order parameter.
A feature of a QCP in metal is characterized by an anomalous electronic state affected by quantum fluctuations.
In the metallic antiferromagnetic (AFM) case, a large number of materials show the AFM QCP by changing control parameters such as pressure, chemical substitution, and magnetic field.
In specific materials, for instance heavy fermion systems or $d$-electron systems such as Fe-pnictides, they can offer a stage to induce exotic superconductivity, and have been extensively studied. 
In the metallic ferromagnetic (FM) case, however, a QCP at zero field as shown in Fig.~1(a) is hardly achieved.\cite{Brando}
A route to avoid a FM QCP is an appearance of the AFM state, as shown in Fig.~1(b).
Actually, many ferromagnets show the alternation of the ground state before the termination of the FM state.\cite{Sidorov, Kotegawa2,Lengyel,Valentin_LaCrGe3,Kaluarachchi,Kaluarachchi2}
More crucial route to prevent a QCP is an appearance of a tricritical point (TCP), above which the second order transition changes to the first order transition, as expressed in Fig.~1(c).
In this case, a metamagnetic transition from a polarized paramagnetic (PM) state to a FM state is induced by applying magnetic field along the easy axis, and the wing structure of the first-order planes is formed.
A QCP emerges at a finite magnetic field, where the first-order metamagnetic transition terminates, but it is not at zero field.
This kind of phase diagram has been investigated theoretically, \cite{Yamada,Belitz,Millis,Binz,Yamaji,Imada} and qualitatively reproduced by experiments in many systems such as MnSi, ZrZn$_2$, UGe$_2$, UCoAl, URhAl and U$_3$P$_4$.\cite{Pfleiderer_MnSi,Uhlarz,Kabeya,Valentin,Kotegawa1,Aoki,Kimura,Shimizu,Araki}
Sr$_3$Ru$_2$O$_7$ is also included in this category.\cite{Wu}
Originally, the change of the order in the FM transition and the emergence of the metamagnetic transition have been discussed to originate in the band structure in the itinerant system, as discussed in UGe$_2$,\cite{Sandeman} but a crucial point is a theoretical suggestion that the FM quantum phase transition is generically first order by considering soft particle-hole excitations about the Fermi surface, which are always present in metals.\cite{Belitz,Belitz2}
If this is correct, a FM QCP is impossible to be realized at zero field.
It is an intriguing challenge to find a FM QCP in actual materials.
As another important theoretical suggestion, it has been proposed that the wing structure is lost by disorder.\cite{Sang}
The QCP at zero field can be obtained by disorder effect, but it is not an intrinsic form of ferromagnets.
An investigation in a clean system is essential.

\begin{figure}[htb]
\centering
\includegraphics[width=\linewidth]{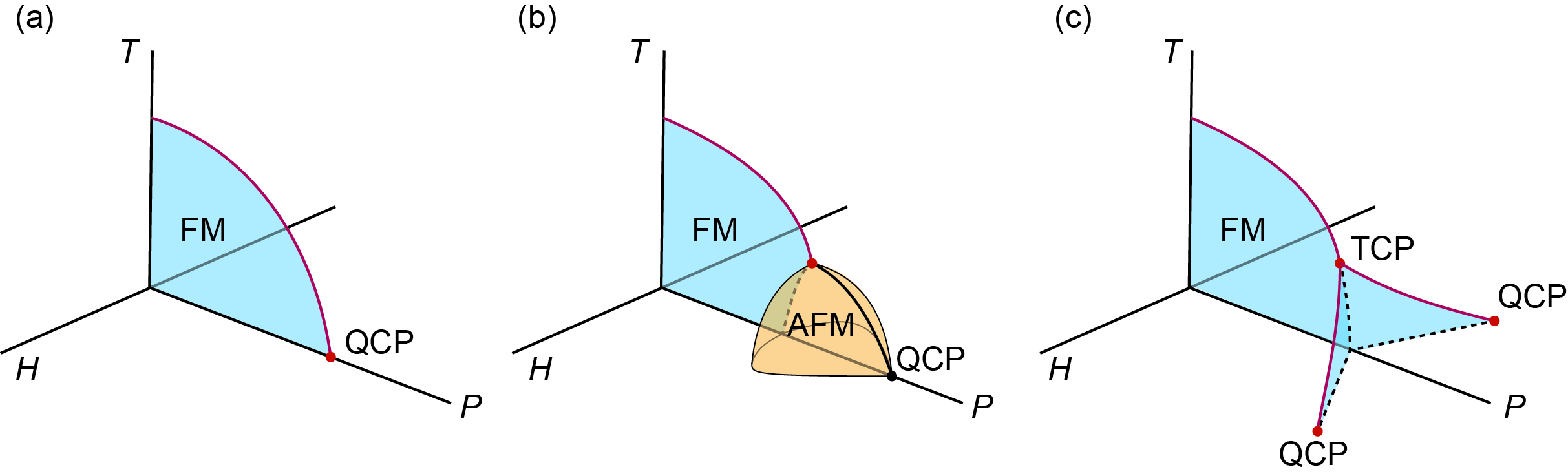}
\caption{(Color online) Schematic phase diagram of metallic ferromagnets for $T$: temperature, $P$: pressure, and $H$: magnetic field.\cite{Brando} (a) The system showing a FM QCP. The second-order FM transition, represented by a red curve, reaches 0 K continuously. (b) The ground state changes from a FM state to a AFM state. (c) The transition changes at a TCP from second order to first order. A metamagnetic transition and the accompanied QCP appear at the finite magnetic fields, but a FM QCP is absent at zero field. }
\end{figure}

A good example being close to a FM QCP is YbNi$_4$P$_2$, which shows extremely low $T_{\rm Curie}=170$ mK.\cite{Krellner}
The FM transition is second order in spite of the low $T_{\rm Curie}$.
Substitution of As for P expands the lattice and suppresses the FM state.
The 10\% doping can induce the FM QCP,\cite{Steppke} but a control by pressure is more desirable to prevent from inducing disorder. 
Another good example is FM superconductor UCoGe with $T_{\rm Curie}=2.5$ K.
The FM state is easily suppressed by pressure, but the order of the FM transition has been controversial.
The bulk measurements have shown the feature of the continuous transition under pressure in UCoGe,\cite{Elena,Slooten,Gael} while the NQR measurement suggests that the FM transition is discontinuous, that is, first order.\cite{Ohta,Manago}
It is an intriguing issue whether this discrepancy is specific to UCoGe or common behavior in metallic ferromagnets, but other examples to be investigated are very few.

\begin{figure}[htb]
\centering
\includegraphics[width=0.7\linewidth]{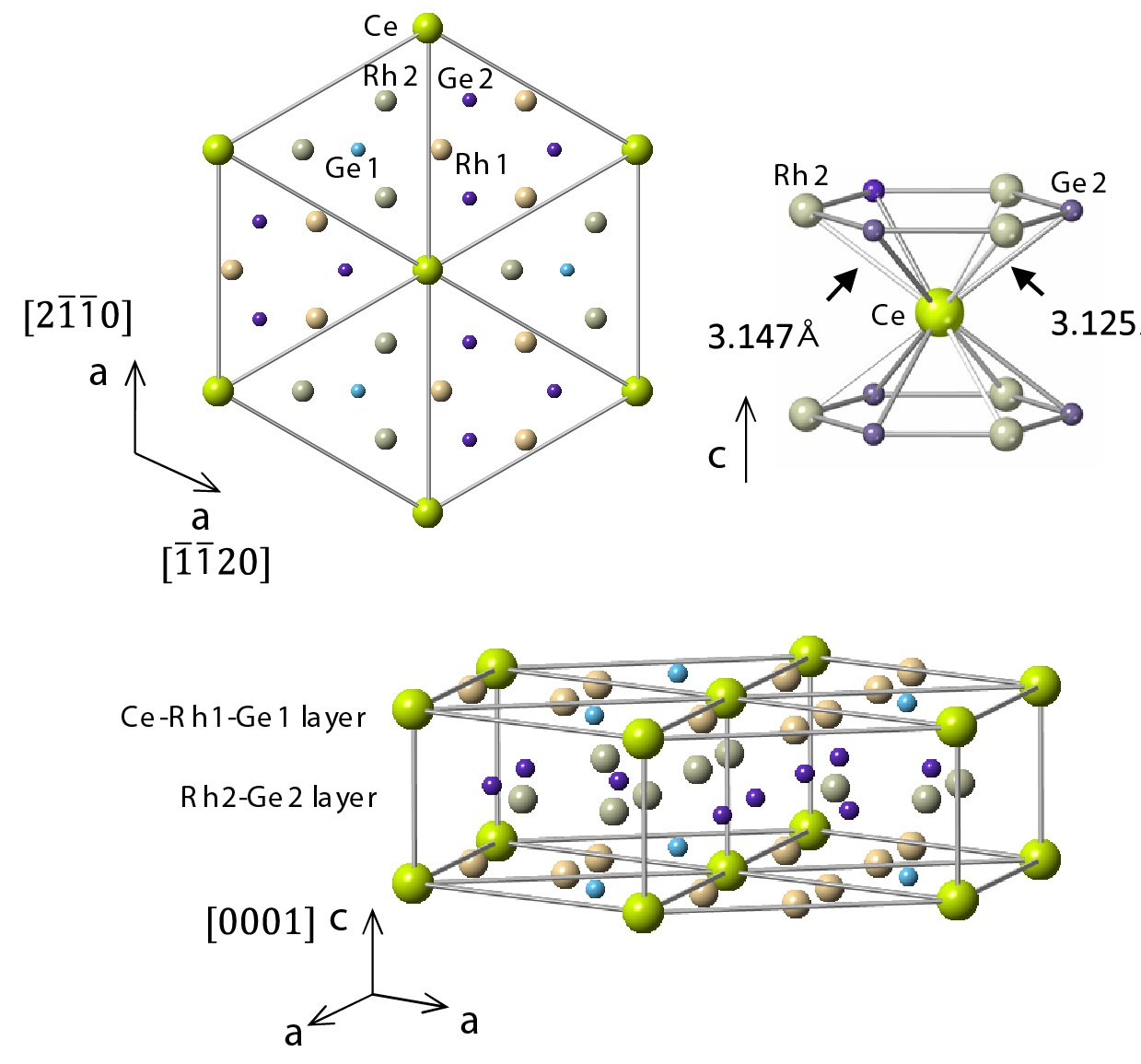}
\caption{(Color online) The crystal structure of the non-centrosymmetric hexagonal CeRh$_6$Ge$_4$, which is composed of Ce-Rh1-Ge1 layer and Rh2-Ge2 layer. The space group is $P$\=6$m$2 (No. 187, $D_{3h}^1$). The Ce ion is surrounded by six Rh2 and six Ge2 ions, and the translational symmetry is realized for all the Ce ions.}
\end{figure}

Recently, CeRh$_6$Ge$_4$ was presented as a novel example, which would be in the vicinity of a FM QCP.\cite{Matsuoka}
It is a stoichiometric Kondo lattice with weak ferromagnetism below $T_{\rm Curie}=2.5$ K.
CeRh$_6$Ge$_4$ crystalizes in non-centrosymmetric hexagonal structure with the space group of $P$\=6$m$2 (No. 187, $D_{3h}^1$). \cite{Voswinkel}
It is similar to UCoAl with $P$\=62$m$ (No. 189, $D_{3h}^3$), which follows the phase diagram shown in Fig.1(c).\cite{Aoki,Kimura}
In CeRh$_6$Ge$_4$, the translational symmetry is realized for all the Ce ions.
It is composed of Ce-Rh1-Ge1 layer and Rh2-Ge2 layer, as shown in Fig.~2.
The nearest Ce-Ce distance is 3.855\AA \ along the $c$ axis, and the Ce-Ge2 distance (3.125 \AA) and Ce-Rh2 distance (3.147 \AA) are shorter. 
The easy axis is perpendicular to the $c$ axis, and the spontaneous magnetization of $\sim0.16$ $\mu_B$/Ce is also induced perpendicular to the $c$ axis below $T_{\rm Curie}$.\cite{Matsuoka2}
The reduced ordered moment and the small entropy of $\sim20$ \% of $R \ln 2$ just above $T_{\rm Curie}$ indicate the strong Kondo effect of this material.\cite{Matsuoka}
The specific heat suggests that the FM transition is of second order; therefore, CeRh$_6$Ge$_4$ is most likely to be located in the vicinity of a FM QCP.
Another notable point is that the residual resistivity ratio (RRR) is more than 30,\cite{Matsuoka,Matsuoka2} ensuring high quality of the crystal.
In this letter, we show the pressure effect of this clean material to reveal how the FM state is suppressed, and discuss the unusual aspect in the band structure.

The single crystal of CeRh$_6$Ge$_4$ was grown by a Bi-flux method.\cite{Voswinkel}
It is needle-shape grown along the $c$ axis, and the typical thickness is about 50 $\mu$m.
The electrical resistivity was measured using a four-probe method, in which the current direction was the $c$ axis.
The electrical contacts of wire were made by a spod-weld method. 
The high pressure was applied by utilizing an indenter-type pressure cell and Daphne7474 as a
pressure-transmitting medium.\cite{Indenter,Murata}
The low temperature was achieved using a dilution refrigerator or a $^3$He cryostat.
Band structure calculation was obtained through a full-potential LAPW (linear augmented plane wave) calculation within the LDA (local density approximation).

\begin{figure}[htb]
\centering
\includegraphics[width=0.9\linewidth]{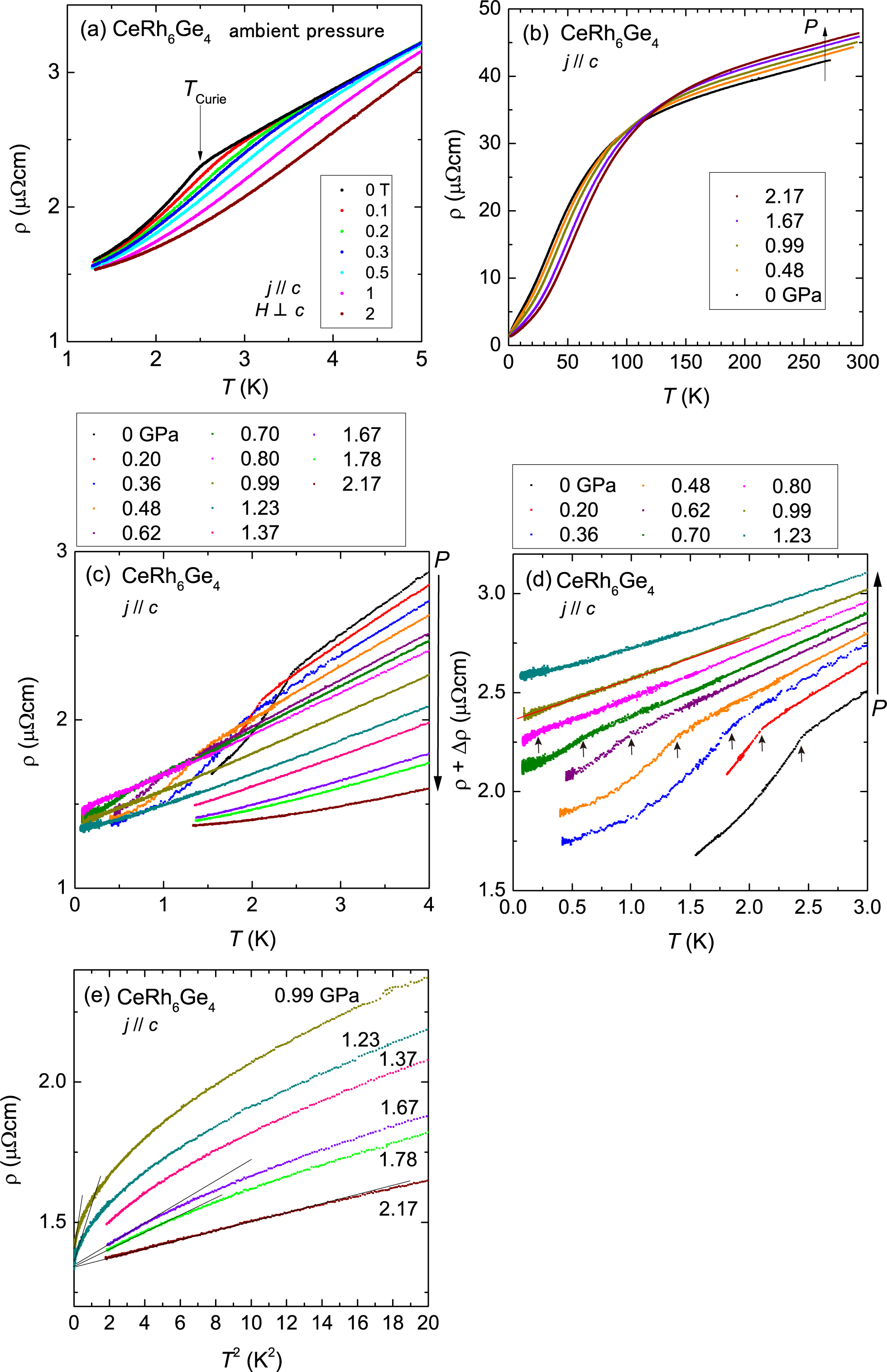}
\caption{(Color online) (a) The electrical resistivity of CeRh$_6$Ge$_4$ at ambient pressure and under magnetic fields. The resistivity at several pressures (b) up to 300 K and (c) in the low-temperature region. (d) The resistivity is shifted to avoid the overlap of the data. Here, the expedient factors $\Delta \rho$, which are proportional to pressure, were added. (e) $\rho$ vs. $T^2$. At 0.99 and 1.23 GPa, the resistivity does not obey the FL form, but the $A$ coefficient was tentatively estimated as shown by the straight lines. Above 1.67 GPa, the FL region starts to appear.}
\end{figure}

Figure 3(a) shows the temperature dependence of the electrical resistivity of CeRh$_6$Ge$_4$ measured 
at ambient pressure and at several magnetic fields along the easy axis.
The FM transition at $T_{\rm Curie}=2.5$ K is clearly seen at zero field, while it easily changes to the crossover under weak magnetic fields.
This is consistent with the interpretation that the ground state of CeRh$_6$Ge$_4$ is the FM state, although it has not been confirmed experimentally whether the FM state is completely collinear or not. 
Figures 3(b-d) show the temperature dependence of the electrical resistivity of CeRh$_6$Ge$_4$ measured at several pressures up to 2.17 GPa.
The low residual resistivity $\rho_0$ of 1.35 $\mu \Omega$cm ensures the high quality of the present single crystal.
The resistivity possesses the shoulder at around 60 K, which is an indication of the Kondo effect, and it moves to higher temperatures as increasing pressure as well as most Ce-based materials.
Figure 3(c) shows the resistivity at low temperatures, and the experimental data are shifted by adding the expedient factors in Fig.~3(d) to avoid the overlap of the data.
The FM transition is clearly observed at $\sim2.5$ K at ambient pressure, and it decreases sensitively and gradually as increasing pressure.
At 0.70 GPa, the transition is still clearly seen at $\sim0.6$ K, and the small anomaly still remains at around $0.2-0.3$ K at 0.80 GPa, as indicated by the arrows.
The anomaly disappears at around 0.99 GPa, where temperature dependence of the resistivity is linear, as indicated by the red line.
This suggests a realization of the non-Fermi liquid (NFL) state in the vicinity of the quantum phase transition between the FM and PM states.
The $T$-linear resistivity differs from a prediction by a SCR theory; $T^{5/3}$ for three-dimensional FM QCP,\cite{Moriya} but is consistent with those observed in YbNi$_4$P$_2$ and UCoGe.\cite{Krellner,Steppke,Elena,Slooten,Gael}
At 1.23 GPa, the feature of the NFL is weakened, and the ground state changes over to the FL state at higher pressures.
The change from the NFL to the FL states can be checked in Fig. 3(e), where the resistivity is plotted as a function of $T^2$.
Linearity, which corresponds to the FL state, is hardly obtained at 0.99 GPa and even at 1.23 GPa, while it comes up above 1.67 GPa at the low-temperature regions.
The slope at low temperatures gives $A$ coefficient in $\rho(T)=\rho_0 + A T^2$, and the $A$ drastically decreases under higher pressures.

\begin{figure}[htb]
\centering
\includegraphics[width=0.75\linewidth]{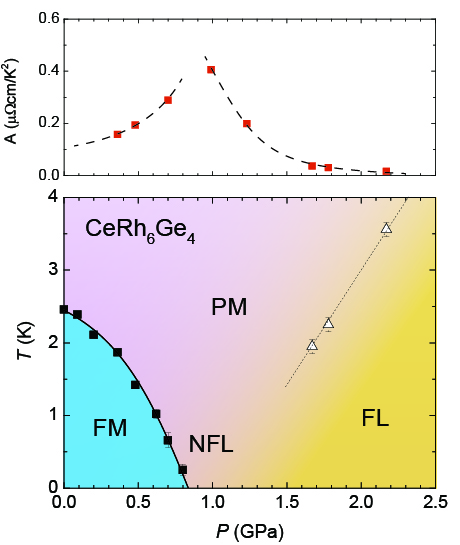}
\caption{(Color online) (bottom): The $P$-$T$ phase diagram of CeRh$_6$Ge$_4$. The closed squares indicate $T_{\rm Curie}$, and the open triangles represent $T_{\rm FL}$, below which the resistivity obeys $T^2$ dependence. The $T_{\rm Curie}$ is suppressed continuously by pressure, and reaches 0 K at $P_C\sim0.85$ GPa. Any clear signature of a first-order transition was not observed in this experiment, and the NFL state is realized at the FM and PM boundary, reminiscent of the presence of the QCP. The FL state is stabilized at higher pressures. (top): pressure dependence of the $A$ coefficient. The $A$'s at 0.99 and 1.23 GPa were tentatively determined, as shown in Fig.~3(e). The divergent behavior appears in the vicinity of $P_C$.}
\end{figure}

The pressure-temperature phase diagram of CeRh$_6$Ge$_4$ is shown in Fig.~4.
The $T_{\rm Curie}$ continuously decreases and seems to reach 0 K at $P_C\sim0.85$ GPa.
The trial fitting gives $T_{\rm Curie} \propto (P_C-P)^{ \beta}$ with $\beta \sim \frac{3}{5}$ (not shown), which is somehow smaller than $\beta =\frac{3}{4}$ predicted for that a three-dimensional clean ferromagnet.\cite{Millis2}
In the experimental resolution, there is no clear signature that the transition changes to first order.
The observation of the NFL state is a strong indication of the critical fluctuations at low temperatures.
The top panel shows the pressure dependence of the $A$ coefficient.
Here, the FL state is not realized at 0.99 and 1.23 GPa but they are determined by a tentative fitting at low temperatures as indicated in Fig.~3(d).
The $A$ shows the divergent behavior, which is reminiscent of the presence of a QCP.

\begin{figure}[htb]
\centering
\includegraphics[width=\linewidth]{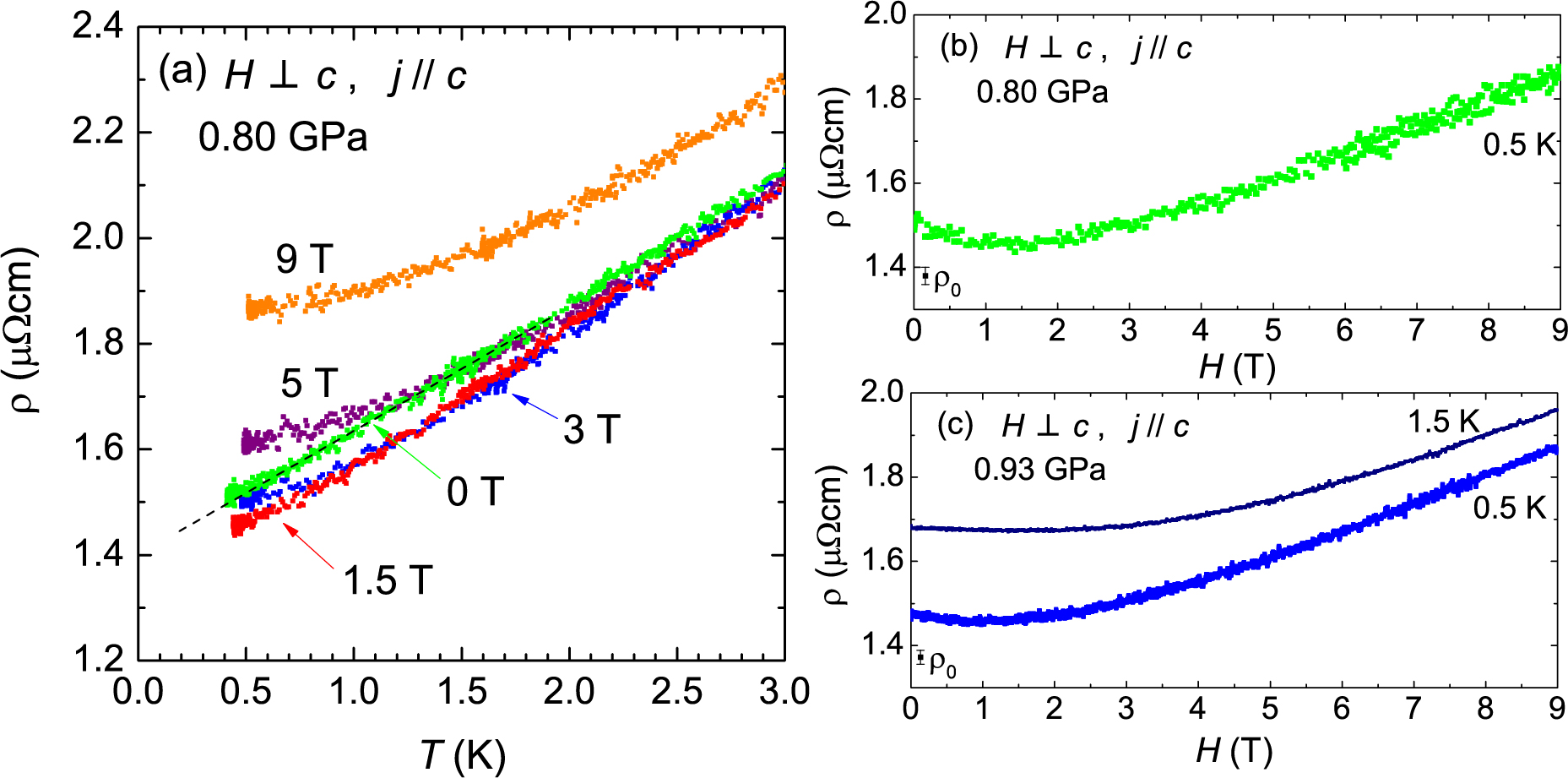}
\caption{(Color online) (a) The resistivity at 0.80 GPa measured at several magnetic fields. The dotted line is drawn for the guide to zero-field data. (b, c) The field dependences of the resistivity at 0.80 and 0.93 GPa. No clear anomaly was observed at 0.5 K within the experimental resolution. Each $\rho_0$ indicates the value estimated from an extrapolation of temperature dependence of $\rho$ at zero field.}
\end{figure}

Figures 5(a-c) show the effects of the magnetic field in the vicinity of $P_C$.
At 0.80 GPa, where $T_{\rm Curie}$ is expected to be close to $\sim0.25$ K, the resistivity at zero field shows the $T$-linear behavior above 0.4 K.
The applied field of 1.5 T suppresses the scattering especially below $\sim2$ K, indicating that the broad crossover occurs at around 2 K.
The NFL behavior at low temperatures is suppressed as increasing magnetic field.
In the field dependence measured at 0.5 K shown in Figs.~5(b) and (c), the resistivity slightly decreases at lower fields due to the suppression of the scattering, and turns to increase towards higher fields due to the occurrence of the transverse magnetoresistance. 
There is no clear signature of the field-induced phase transition corresponding to the case shown in Fig.~1(c).
If a system has a metamagnetic transition, which is relevant with a presence of a TCP, the corresponding anomaly is often detected in resistivity,\cite{Kotegawa1,Aoki,Kimura} but they are negligible in CeRh$_6$Ge$_4$ within the experimental resolution.
This suggests that a TCP in CeRh$_6$Ge$_4$ is absent or located at least below $\sim0.5$ K even if it exists.
In most of the clean ferromagnets, the TCP appears above 5 K;\cite{Brando} therefore, the electronic state in CeRh$_6$Ge$_4$ is quite unusual.

\begin{figure}[htb]
\centering
\includegraphics[width=\linewidth]{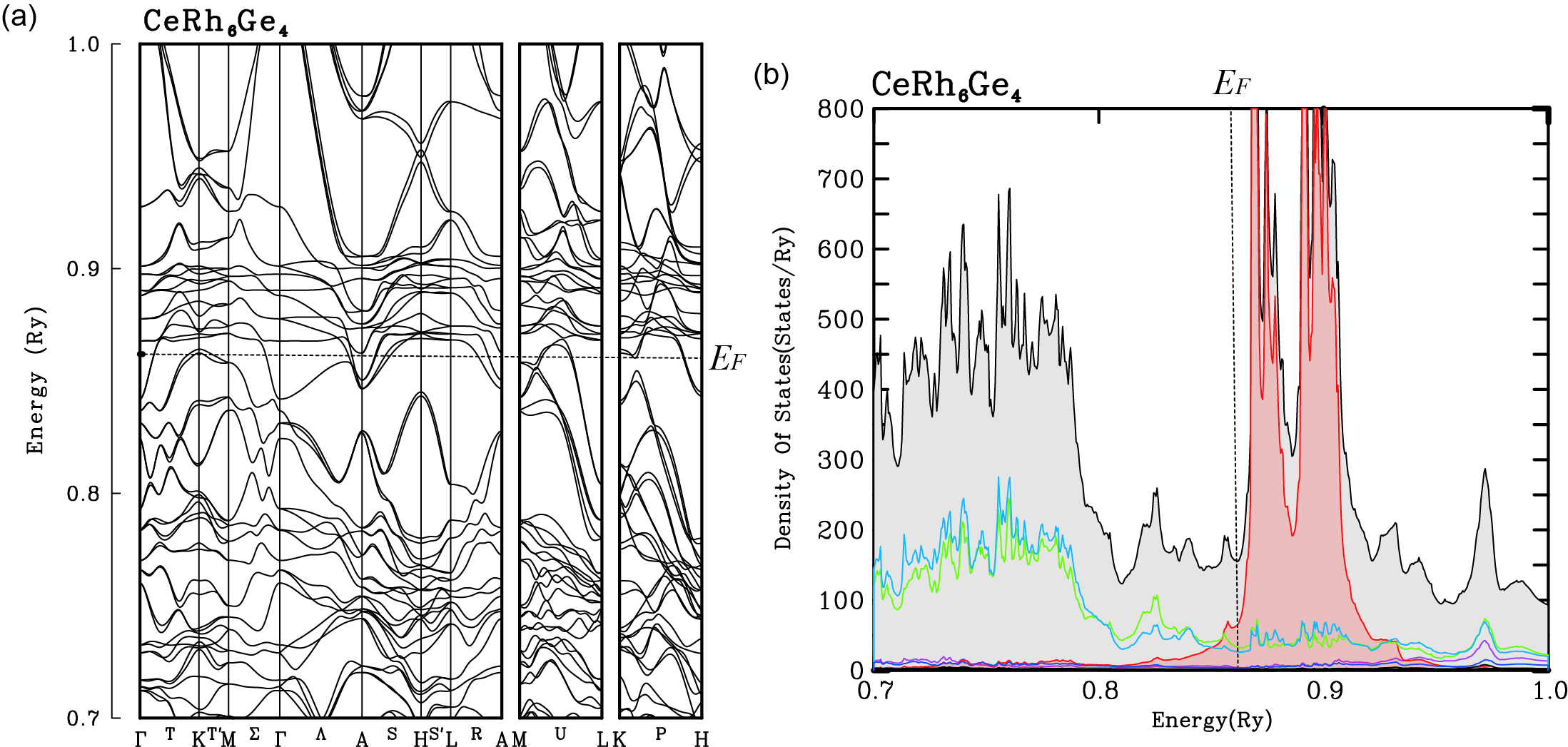}
\ \vspace{3ex}
\includegraphics[width=0.8\linewidth]{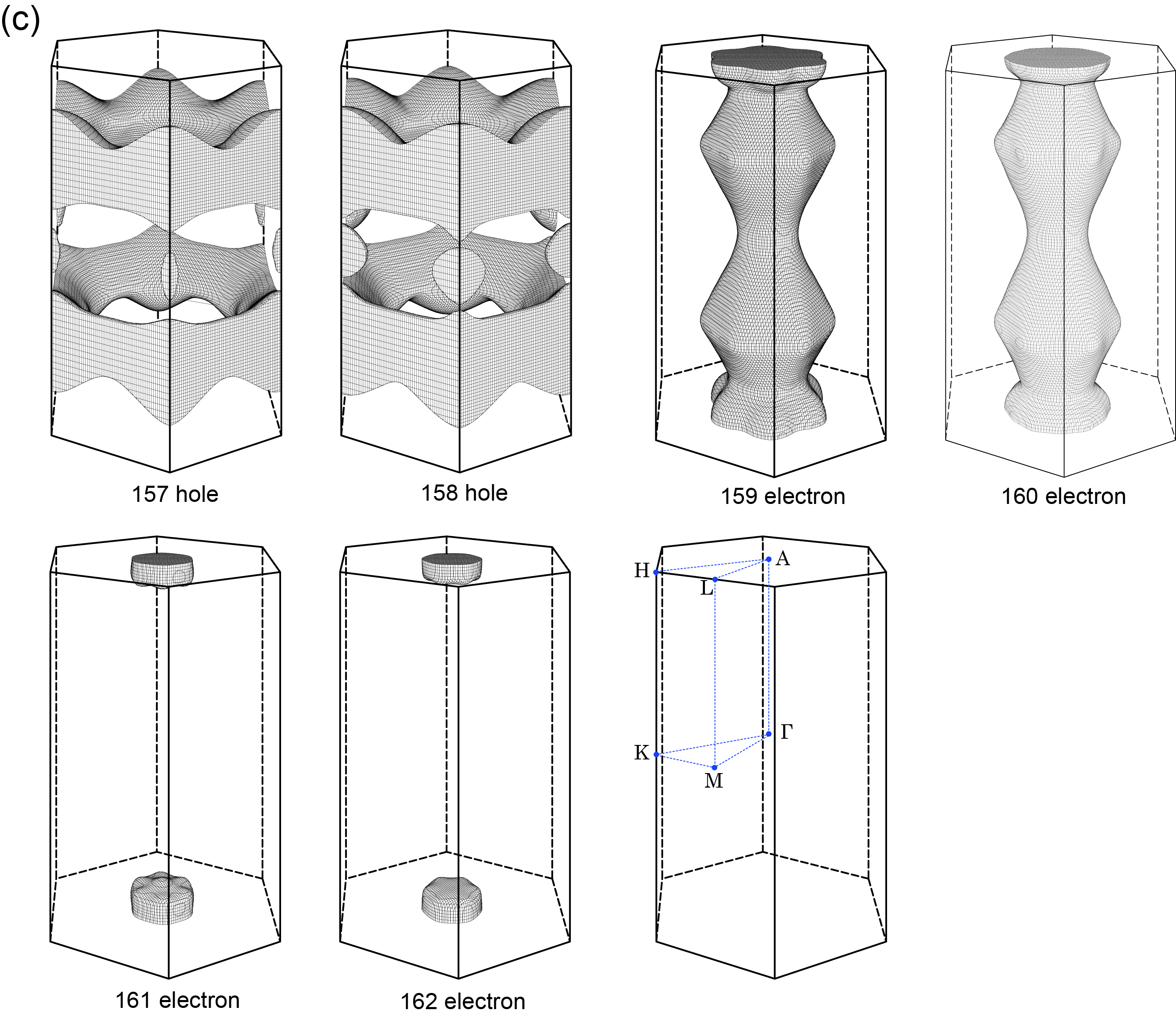}
\caption{(Color online) (a) Calculated band structure in the PM state for CeRh$_6$Ge$_4$. (b) Density of states near the Fermi energy. The colored curves indicate the partial DOS; Ce-$4f$ (red), Rh1-$4d$ (green), Rh2-$4d$ (blue), Ce-$5d$ (purple). The black curve shows the total DOS. (c) The Fermi surfaces of CeRh$_6$Ge$_4$.}
\end{figure}

In YbNi$_4$P$_2$, one-dimensionality of the Fermi surface has been proposed to be a key factor to avoid a first-order phase transition,\cite{Krellner,Steppke} and it still opens to argument.
To examine the electronic state of CeRh$_6$Ge$_4$, we performed the band structure calculation utilizing the lattice parameters obtained in Ref.~35.
Figure 6(a) and (b) show the energy dispersion and the density of states (DOS) near the Fermi level for the PM state.
The partial DOS originating in Ce-$4f$ orbits forms the two-peak structure above $E_F$.
The $E_F$ is located at the tail of the peak structure of Ce-$4f$ orbits, or the local minima in the total DOS, giving the relatively low and gently-sloping DOS at $E_F$.
This is in contrast to the typical LDA calculation for Ce-based heavy fermion compounds, where $E_F$ is located at the shoulder of the peak structure,\cite{Sheikin,Ishikawa} even in a nonmagnetic system.\cite{Suzuki}
It should be noted that the valence of Ce in CeRh$_6$Ge$_4$ is suggested to be close to 3+, because of the sufficiently occupied electron number in the tail.
The relatively low DOS must be relevant with the weak magnetism in CeRh$_6$Ge$_4$, and the absence of the steep structure in the DOS in the vicinity of $E_F$ can be an important factor to avoid a clear discontinues FM transition. 
The two peaks originating in Ce-$4f$ orbits are separated well by the atomic spin-orbit coupling, indicating weak hybridization effect, whereas the significant broadened tail is brought by the hybridization.
These two features suggest that only specific orbital in Ce-$4f$ has the strong hybridization with the conduction electrons, probably giving a key ingredient to induce the unusual electronic state in CeRh$_6$Ge$_4$.
The calculated Fermi surfaces are shown in Fig, 6(c), where two sets of the Fermi surfaces appear because of lack of the space inversion symmetry.
Two-dimensional feature is seen at the 159 and 160 electron sheets, and a similarity to the Fermi surface of YbNi$_4$P$_2$ is not clear.\cite{Krellner}
Experimental determination of the Fermi surfaces is an important issue to clarify the origin inducing the unusual electronic state in CeRh$_6$Ge$_4$.

In conclusion, we established the $P$-$T$ phase diagram of the novel weak ferromagnet CeRh$_6$Ge$_4$.
The $T_{\rm Curie}$ of 2.5 K is easily suppressed by a low pressure of $P_C\sim0.85$ GPa.
The continuous disappearance of the FM state and the obvious NFL behavior suggest the presence of the FM QCP at zero field or at least in the immediate vicinity of zero field.
CeRh$_6$Ge$_4$ behaves as a material showing a FM QCP.
In UCoGe, however, the FM transition has been suggested to be microscopically discontinuous to the contrary to the continuous behavior in other bulk measurements.
It is interesting to check the electronic state in the vicinity of $P_C$ for CeRh$_6$Ge$_4$ from the microscopic point of view.
This would be a crucial issue to clarify what happens at the limit where a TCP goes to zero.
Unfortunately, superconductivity did not appear in CeRh$_6$Ge$_4$, but this may be a benefit for the research of the pure quantum FM transition.
Another notable point is the interplay between the matter of a quantum FM transition and the band structure.
The band structure calculation suggests the peculiar electronic state in CeRh$_6$Ge$_4$.
CeRh$_6$Ge$_4$ is a key material to know whether a FM QCP can be realized in actual material, and also how to achieve it.

\section*{Acknowledgements}
This work was supported by JSPS KAKENHI Grant Number JP15H05882,
JP15H05885, JP18H04320, and 18H04321 (J-Physics), 15H03689.and 15H05745.

\end{document}